%Paper: gr-qc/9411011
%From: rendall@ihes.fr (Alan Rendall)
%Date: Fri, 4 Nov 1994 15:56:50 +0100 (MET)

\noindent
\magnification=1200
\def\tr#1{{\rm tr #1}}
\def\d{\partial}
\def\f#1#2{{\textstyle{#1\over #2}}}

\def\next{\hfil\break\noindent}
\def\R{{\bf R}}

\font\title=cmbx12
\hfuzz=2 pt

{\title
\centerline{Crushing singularities in spacetimes with}
\centerline{spherical, plane and hyperbolic symmetry}}

\vskip 1cm

\noindent
Alan D. Rendall
\next
Max-Planck-Institut f\"ur Astrophysik
\next
Karl-Schwarzschildstr. 1
\next
Postfach 1523
\next
85740 Garching
\next
Germany

\vskip 10pt\noindent
and

\vskip 10pt\noindent
Institut des Hautes Etudes Scientifiques
\next
35 Route de Chartres
\next
91440 Bures sur Yvette
\next
France\footnote{*}{Present address}

\vskip 1cm\noindent
{\bf Abstract} It is shown that the initial singularities in spatially
compact spacetimes with spherical, plane or hyperbolic symmetry
admitting a compact constant mean curvature hypersurface are
crushing singularities when the matter content of spacetime is
described by the Vlasov equation (collisionless matter) or the wave
equation (massless scalar field). In the spherically symmetric case
it is further shown that if the spacetime admits a maximal slice
then there are crushing singularities both in the past and in the
future. The essential properties of the matter models chosen are
that their energy-momentum tensors satisfy certain inequalities
and that they do not develop singularities in a given regular
background spacetime.

\vskip 2cm\noindent
{\bf 1. Introduction}

The nature of singularities in general solutions of Einstein's
equations is still a matter about which very little is known.
The best information on this which has been obtained up to now
concerns the singularities in special classes of solutions defined by
symmetry assumptions. The hope is that the insights obtained in
solving these restricted problems will allow the symmetry
assumptions to be progressively relaxed and so the study of
singularities in solutions of the Einstein equations with various
symmetries can be seen as a systematic approach to the general
problem of understanding spacetime singularities. The following is
intended as a contribution in this direction.

At the moment the study of singularities in solutions of the vacuum
Einstein equations is significantly further advanced than in the case
of the Einstein equations coupled to matter.  This paper is concerned
mainly with non-vacuum spacetimes and helps to redress the balance a
little. The spacetimes treated have spherical, plane or hyperbolic
symmetry and a compact Cauchy hypersurface. (The reader wishing to
gain some intuition for this class of spacetimes is referred to
Appendix 2, where the vacuum solutions are determined explicitly.)
Some of the results are for general matter models which are only
restricted by some inequalities on the components of the
energy-momentum tensor. However the main results require a matter
model which is well-behaved in a certain sense. Roughly speaking, the
matter fields should not develop singularities in a given regular
spacetime. These results are worked out for two particular matter
models, namely collisionless matter and the massless scalar field
(Theorems 3.2 and 3.3 respectively).

There are some differences between the results for the different
symmetry classes. For plane symmetry it is shown that,
if the spacetime is the maximal globally hyperbolic development
of constant mean curvature (CMC) initial data on a compact spacelike
hypersurface, there exists a foliation of compact hypersurfaces of
constant mean curvature which covers either the past or the future
of the initial hypersurface. If the mean curvature of the initial
hypersurface is $H_0$ then the range of the mean curvature of this
foliation contains the interval $(-\infty, H_0]$ or $[H_0,\infty)$
respectively. The analogous result in the case of hyperbolic
symmetry is proved under the additional assumption that the mass
function, defined by equation (2.15), is positive on the initial
hypersurface. In the spherically symmetric case the results are
as follows. If the initial hypersurface is maximal (i.e. its
mean curvature is zero) then it is shown that the spacetime can be
covered by a CMC foliation where this time the range of the mean
curvature is the whole real line. This should probably be true
whatever the mean curvature of the initial hypersurface but all that
could be shown is that there is a foliation by compact CMC hypersurfaces
whose range includes all positive real numbers or all negative real
numbers and this foliation will in general not cover the whole
spacetime. It does however cover the part of the spacetime either to
the past or to the future of the initial hypersurface.  All these
results are dependent on the choice of a well-behaved matter model, as
indicated above. In the spherically symmetric case they are also
dependent on the assumption that the topology of the Cauchy
hypersurface is $S^2\times S^1$ so that there is no centre of
symmetry. The differences between the various cases are related to the
fact that while an initially expanding solution with plane or hyperbolic
symmetry can be expected to expand for ever a spherically symmetric
solution can be expected to recollapse.

In [6] and [7] conjectures were formulated concerning the existence
of CMC foliations. The results just discussed are closely related
to conjecture 2.3 of [7] and conjecture C2 of [6] for the class of
spacetimes considered here. They show in particular that any of these
spacetimes has a crushing singularity in at least one time direction
(and in both time directions if it contains a maximal hypersurface).
Recall that a crushing singularity is one where there exists a
foliation on a neighbourhood of the singularity whose mean curvature
tends uniformly to infinity as the singularity is approached[7].
A difficulty in studying spacetime singularities which may exist at the
boundary of the maximal Cauchy development is that the definition of
the maximal Cauchy development is so abstract. (Zorn's Lemma is used
to show its existence.) Having a geometrically defined global time
coordinate helps to make it more concrete. It allows the strong
cosmic censorship hypothesis to be reformulated as a question on
the global existence and asymptotic behaviour of solutions of a system
of partial differential equations[6]. In this sense proving the
existence of global CMC foliations in a given class of spacetimes
is a first step towards proving strong cosmic censorship in
this class. It also guarantees that a numerical calculation done
using a CMC slicing is in principle capable of covering the whole
spacetime. (The possibility of using a CMC slicing for numerical
studies of cosmological spacetimes has been discussed in [9].)

The most general results on the existence of CMC foliations in
any class of spacetimes with compact Cauchy hypersurfaces are due
to Isenberg and Moncrief[10]. They showed that Gowdy spacetimes
(which are vacuum spacetimes with $U(1)\times U(1)$ symmetry)
on the torus can be foliated globally by constant mean curvature
slices. For spacetimes with matter less is known. In [13] it was
shown that spatially homogeneous spacetimes with matter described
by the Vlasov equation can be foliated by CMC hypersurfaces with
the mean curvature ranging either over the whole interval $(-\infty,
0)$ or the whole real line, depending on the symmetry type.  In
[15] a similar result was obtained for perfect fluids and for
non-interacting mixtures of perfect fluids.

There is another motivation for the work reported in this paper which
has not yet been mentioned. When investigating the global structure
of spacetimes it seems essential to introduce some auxiliary elements
such as coordinates in order to \lq find one's way in spacetime\rq.
If the spacetimes being studied have symmetry properties then it
may be advantageous to use elements which exist due to the symmetry.
The problem with this, from the point of view of the general programme
outlined above, is that this approach is less likely to produce
techniques which can be used beyond the given symmetry type. The
definition of a CMC hypersurface does not depend on any symmetry
assumptions and there is at least a chance that global CMC foliations
will exist in rather general spacetimes. One aim of this work was to
see whether it was possible to prove something about the properties
of singularities working with a CMC foliation from the beginning.
This turned out to be the case.

It is plausible that using coordinates adapted to a special symmetry
type could be more efficient for proving sharper results than CMC
slicing. This is confirmed by the fact that more precise information
about the nature of the singularities in the spacetimes considered in
this paper has been obtained in certain cases using other
coordinate systems. In [14] the case of a plane symmetric scalar
field was studied. It was shown that the initial singularity is
a curvature singularity and a velocity-dominated singularity in
that case. These results provide a model for what one would like
to do more generally. For the case of collisionless matter Rein[12]
has shown that under certain assumptions on the initial data
the initial singularity is a curvature singularity and possesses
some attributes of a velocity-dominated singularity.

\vskip .5cm\noindent
{\bf 2. Analysis of the field equations}

The spacetimes of interest here are defined on manifolds of the form
$M=\R\times S^1\times F$, where $F$ is a compact orientable surface.
Let $p$ denote the projection of the universal cover $\tilde F$
onto $F$. Let $g_{\alpha\beta}$ be a globally hyperbolic metric on $M$
for which each submanifold $\{t\}\times S^1\times F$ is a Cauchy
hypersurface.  The spacetime $(M,g_{\alpha\beta})$ is
called spherically symmetric if $F=S^2$ and the transformations of $M$
induced by the standard action of $SO(3)$ on $S^2$ are isometries
which leave any matter fields invariant. In the case $F=T^2$ the
projection $p$ induces a projection $\hat p$ of $\tilde M=\R\times
S^1\times\R^2$ onto $M$. Let $\tilde g_{\alpha\beta}$ be the pull-back
of $g_{\alpha\beta}$ by $\hat p$. The spacetime is called plane
symmetric if $F=T^2$ and if the tranformations of $\tilde M$ induced
by the standard action of the Euclidean group $E_2$ on $\R^2$ are
isometries which leave (the pull-backs of) any matter fields
invariant. In the case where the genus of $F$ is greater than
one $\tilde F$ can be identified with the hyperbolic plane $H^2$.
A projection $\hat p$ can be defined as in the plane symmetric case.
The spacetime is said to have hyperbolic symmetry if the genus of
$F$ is greater than one and the transformations of $\tilde M=\R\times S^1
\times H^2$ induced by the action of the connected component of the
identity of the isometry group of the hyperbolic plane are isometries
which leave any matter fields invariant. In the spherically symmetric case
$SO(3)$ acts on $S^1\times S^2$ without fixed points. In other words there
are no centres. The group $SO(3)$ also acts on $S^3$ with fixed points and
this leads to a different class of spherically symmetric
spacetimes. This other class will not be considered in this paper and
so for clarity the spherically symmetric spacetimes considered here
will be referred to as spherically symmetric spacetimes without
centre. It will be convenient to refer to spherically symmetric
spacetimes without centre and spacetimes with plane and hyperbolic
symmetry collectively as surface symmetric spacetimes. The surfaces
diffeomorphic to $F$ which are defined by the product decomposition will
be referred to as surfaces of symmetry.

Now the Einstein equations for surface symmetric spacetimes
will be analysed in a certain coordinate system. In order to ensure the
existence of such a coordinate system it will be assumed that the spacetime
possesses a Cauchy hypersurface of constant mean curvature which is
symmetric, in the sense that it is a union of surfaces of symmetry.

\noindent
{\bf Lemma 2.1} Let $(M,g)$ be a non-flat surface symmetric spacetime
having a symmetric constant mean curvature Cauchy
hypersurface and satisfying the dominant energy and non-negative
pressures conditions.
Then in a neighbourhood of any point there exist local coordinates
adapted to the product decomposition $\R\times S^1\times F$ such that
the metric takes the form
$$-\alpha^2 dt^2+A^2[(dx+\beta dt)^2+a^2 d\Sigma^2].\eqno(2.1)$$
The functions $\alpha$, $\beta$ and $A$ depend
on $t$ and $x$, $a$ depends only on $t$ and the metric $d\Sigma^2$ has
constant curvature $\epsilon$. It may be assumed without loss of
generality that $\epsilon=1$ for spherical symmetry, $\epsilon=0$
for plane symmetry and $\epsilon=-1$ for hyperbolic symmetry. The time
coordinate $t$ may be chosen so that the hypersurface where $t$ has a
given value has constant mean curvature equal to that value.

\noindent
{\bf Proof} See Appendix 1.

\vskip 10pt
The neighbourhood in this lemma may be chosen to be the product
$I_1\times I_2\times U$, where $I_1$ and $I_2$ are intervals and
$U$ is an open subset of $F$. The interval $I_1$ will be denoted
by $(t_1,t_2)$ in the following. By a slight extension of the usual
notion of coordinates $I_2$ can be taken to be the closed interval
$[0,2\pi]$, where it is understood that $x=0$ and $x=2\pi$ are to
be identified. The functions $\alpha$, $\beta$ and $A$ are assumed
to be functions which have $C^\infty$ extensions to functions
which are $2\pi$-periodic in $x$. The coordinates can be chosen
so that $\int_0^{2\pi}\beta(t,x) dx=0$ for each $t$. The second
fundamental form of each hypersurface $t={\rm const.}$ can be written as
$$A^2(K dx^2-\f12(K-t)d\Sigma^2)\eqno(2.2)$$
for a function $K(t,x)$ which has the same regularity properties
as those demanded of $\alpha$, $\beta$ and $A$ above. The Einstein
equations will now be written out for the metric (2.1), making use
of the variable $K$ defined by (2.2).

$$\eqalignno{
&(A^{1/2})''=-\f18A^{5/2}[\f32(K-\f13t)^2-\f23t^2+16\pi\rho]
+\f14\epsilon A^{-1/2}a^{-2}&(2.3)              \cr
&\alpha''+A^{-1}A'\alpha'=\alpha A^2[\f32(K-\f13t)^2+\f13 t^2+4\pi(\rho
+\tr S)]-A^2&(2.4)                              \cr
&K'+3A^{-1}A'K-A^{-1}A't=8\pi jA&(2.5)          \cr
&\beta'=-a^{-1}\d_t a+\f12\alpha(3K-t)&(2.6)    \cr
&\d_t a=a[-\beta'+\f12\alpha(3K-t)]&(2.7)             \cr
&\d_t A=-\alpha KA+(\beta A)'&(2.8)             \cr
&\d_t K=\beta K'-A^{-2}\alpha''+A^{-3}A'\alpha'& \cr
&\qquad+\alpha[-2A^{-3}A''
+2A^{-4}{A'}^2+Kt-8\pi S^1_1+4\pi\tr S-4\pi\rho]
&(2.9)}
$$
The primes here denote derivatives with respect to $x$.
The equation (2.3) is the Hamiltonian constraint while (2.5) is
the momentum constraint. The constant mean curvature condition
leads to the lapse equation (2.4). Equation (2.6) is a consequence
of the choice of spatial coordinate condition. Equations (2.7)
and (2.8) come from the definition of the second fundamental form
and (2.9) is the one independent Einstein evolution equation
which exists in this situation. To give the definition of the matter
quantities occurring in (2.9) it is convenient to introduce a
(locally defined) orthonormal frame. Let $e_0$ be the future-pointing
unit normal to the hypersurfaces of constant $t$. Let $e_1$ be a
unit vector tangent to these hypersurfaces which is normal to the
surfaces of constant $t$ and $x$. Complete $e_0$ and $e_1$ to an
orthonormal frame by adding vectors $e_2$ and $e_3$. Then
$\rho=T_{\alpha\beta}e_0^\alpha e_0^\beta$, $j=-T_{\alpha\beta}
e_0^\alpha e_1^\beta$, $S^1_1=T_{\alpha\beta}e_1^\alpha e_1^\beta$
and $\tr S=T_{\alpha\beta}(e_1^\alpha e_1^\beta+e_2^\alpha e_2^\beta
+e_3^\alpha e_3^\beta)$.

Malec and \'O Murchadha [11] have written the constraints in an
alternative form which turns out to be very useful in certain
circumstances. (In fact they only consider the spherically
symmetric case but the extension to plane and hyperbolic symmetry
is straightforward.) They use as fundamental variables the
expansions $\theta$ and $\theta'$ of the families of null
geodesics orthogonal to the orbits. (The prime in $\theta'$ is
not a derivative.) In terms of the above coordinates these are
given explicitly by
$$\eqalign{
&\theta=2A^{-2}A'+K-t                     \cr
&\theta'=2A^{-2}A'-K+t}
\eqno(2.10)$$
The sign conventions for $\theta$ and $\theta'$ are those of [11] but
the sign convention for the second fundamental form is the opposite
of that used in [11].
The normalization of $\theta$ and $\theta'$ depends on a choice of
Cauchy hypersurface. Equation (2.10) applies to the case where the
hypersurface chosen is a level surface of the coordinate $t$. However
the analysis of [11] applies to any Cauchy hypersurface $S$ compatible
with the product decompostion of $M$ in the sense that it is a union
of surfaces of symmetry. The mean curvature of a Cauchy hypersurface
of this kind will be denoted by $H$ and is in general not constant.
Let $l$ be a proper length parameter along a curve in $S$ orthogonal
to the surfaces of symmetry. When expressed in
terms of $\theta$, $\theta'$ and $l$ the constraints (2.3) and (2.5)
become:
$$\eqalign{
&\d_l\theta=-8\pi(\rho-j)-\f34\theta^2-\theta H+\epsilon (aA)^{-2}
\cr
&\d_l\theta'=-8\pi(\rho+j)-\f34\theta^{'2}+\theta'H+\epsilon (aA)^{-2}
}\eqno(2.11)$$
For the following discussion it is useful to introduce the area radius
$r=aA$. The idea of [11] is to use the equations (2.11) to obtain bounds
for the quantities $r\theta$ and $r\theta'$. It is assumed that the
dominant energy condition holds so that $|j|\le\rho$. From (2.11)
$$\eqalign{
&\d_l(r\theta)=-8\pi r(\rho-j)-{1\over 4r}(\theta^2 r^2-4\epsilon
+4\theta Hr^2+\theta r(\theta r-\theta' r))              \cr
&\d_l(r\theta')=-8\pi r(\rho+j)-{1\over 4r}(\theta^{'2} r^2-4\epsilon
-4\theta' Hr^2+\theta'r(\theta'r-\theta r))}\eqno(2.12)$$
Consider now one particular Cauchy hypersurface $S$ which is a
union of surfaces of symmetry. Denote the maximum
value attained by $r\theta$ and $r\theta'$ on this hypersurface by
$M_+$ and the minimum by $M_-$. Let $x_0$ be a point
where $M_+$ is attained and suppose
without loss of generality that $\theta(x_0)\ge \theta'(x_0)$.
Since $x_0$ is a critical point of $r\theta$ it follows from (2.12)
that at that point
$$\theta^2 r^2+(4Hr)(\theta r)-4\epsilon\le 0\eqno(2.13)$$
Working out the roots of the corresponding quadratic equation then
shows that
$$-2(|Hr|+\sqrt{\epsilon+H^2 r^2})\le M_+
\le 2(|Hr|+\sqrt{\epsilon+H^2 r^2})\eqno(2.14)$$
The same inequality holds for $M_-$.
For each symmetry type this implies that $r\theta$ and $r\theta'$
can be bounded in terms of $H$ and $r$. An important
quantity in the following is the mass function $m$ which is defined
by
$$\epsilon-2m/r=\f14 r^2\theta\theta'\eqno(2.15)$$
The boundedness result just obtained shows that the following holds:

\noindent
{\bf Lemma 2.2} Let $(M,g)$ be a surface symmetric spacetime
which is foliated by compact CMC hypersurfaces with the mean
curvature varying in a finite interval $(t_1,t_2)$. If the dominant
energy condition holds and $r$ is bounded then $2m/r$ is bounded.

\vskip 10pt
There is another formulation of the equations which is also useful.
This is based on the fact that the field equations can be written
as equations on the two dimensional manifold of symmetry orbits.
In the following lower case Latin indices take the values 0 and 1
and are used to express tensor equations on this quotient manifold.
In particular $g_{ab}$ and $T_{ab}$ denote the tensors on the
quotient manifold naturally related to the spacetime metric and the
energy-momentum tensor. The equations of importance in the following
are
$$\eqalignno{
\nabla_a\nabla_b r&={m\over r^2}g_{ab}-4\pi r(T_{ab}-\tr T g_{ab})
&(2.16)   \cr
\nabla_a m&=4\pi r^2(T_{ab}-\tr T g_{ab})\nabla^br&(2.17)}$$
Note that these equations do not contain $\epsilon$ explicitly.
The expression for $m$ in this formulation is
$$m=\f12 r(\epsilon-\nabla^a r\nabla_a r)\eqno(2.18)$$
The following is a generalization of an argument of Burnett[2].

\noindent
{\bf Lemma 2.3} Let $(M,g)$ be a surface symmetric spacetime
and let $S$ be a compact Cauchy hypersurface of the form
$\bar S\times F$ for a curve $\bar S$ in $\R\times S^1$. Let
$r_{\rm min}$ and $r_{\rm max}$ denote the minimum and maximum
values of $r$ on $S$ respectively. Define $R_\epsilon$ to be
$r_{\rm min}$, $0$ or $-r_{\rm max}$ for $\epsilon$ equal to
$1$, $0$ or $-1$ respectively. Then $2m\ge R_\epsilon$ on $S$.

\noindent
{\bf Proof} Let $U$ be the set of $x\in S$ where $2m<R_\epsilon$.
Then $\nabla^a r$ is spacelike on $U$. Let $s_a$ be the projection
of the gradient of $r$ onto $S$. Then
$$s^a\nabla_a m=(T_{ab}-\tr T g_{ab})s^a\nabla^b r\eqno(2.19)$$
If the dominant energy condition holds then the right hand side
of (2.19) is non-negative. Thus on $U$ the mass $m$ increases in
the direction in which $r$ increases.  Now the restriction of $r$ to
$S$ cannot have a stationary point in $U$. This means in particular
that $S\backslash U$ is non-empty. If $x_0$ is a point of
$U$ then moving from $x_0$ in the direction of increasing $r$ leads
to an increase in $m$. Eventually a point of the boundary of $U$
must be reached. At that point $m=R_\epsilon/2$. Thus
$2m(x_0)\le R_\epsilon$. Moving away from $x_0$ in the opposite direction
gives the reverse inequality. Hence $2m=R_\epsilon$ on $U$. However this
contradicts the definition of $U$ unless $U$ is empty.

\vskip 10pt
In the spherical case Lemma 2.3 implies that the minimum of $2m$ on $S$
is greater than or equal to the minimum of $r$ there. For plane
symmetry it implies that the mass is non-negative. The latter
conclusion can be strengthened using (2.11) to get a kind of
positive mass theorem.

\noindent
{\bf Lemma 2.4} Let $(M,g)$ be a plane symmetric spacetime which
satisfies the dominant energy condition. Then if $m=0$ at
any point the spacetime is flat.

\noindent
{\bf Proof} Consider any compact Cauchy hypersurface $S$ of the form
$\bar S\times F$. If $m$ vanishes
at some point $x_0$ of $S$ then (for $\epsilon=0$) either $\theta$
or $\theta'$ must vanish there. Suppose without loss of generality
that it is $\theta$. Equation (2.11) gives
$$\d_l\theta=-(H+\f34\theta)\theta-8\pi (\rho-j)\eqno(2.20)$$
This equation is similar to one which arises in a similar
context for Gowdy spacetimes[5] and can be treated in exactly
the same way. In fact if $l$ is an arc length parameter which
is zero at $x_0$ then the solution of (2.20) is
$$\theta(l)=-8\pi\int_0^l(\rho-j)(u)\exp\left[\int_u^l(-H-\f34\theta)(v)
dv\right] du\eqno(2.21)$$
{}From this formula it is clear that $\theta$ is everywhere
non-positive and that it can only become zero for some positive
$l$ if $\rho-j$ vanishes identically on the interval $[0,l]$.
In that case $\theta$ also vanishes on that interval. Since
$\theta(l)$ is a periodic function it follows that it must be
identically zero and that $\rho=j$ everywhere on $S$. The mass is
also zero on $S$. When $\theta$ is zero the rate of change of $r$
along $\bar S$ is given by $\theta'$. Since the restriction of $r$
to $\bar S$ must have a critical point somewhere, $\theta'$ must
vanish somewhere. Applying to
$\theta'$ the argument previously applied to $\theta$ shows
that $\theta'=0$ and that $\rho=j=0$. By the dominant energy
condition $\rho=0$ implies the vanishing of the whole
energy-momentum tensor on $S$. When the dominant energy condition
holds the vanishing of the energy-momentum tensor on a Cauchy
hypersurface implies that it vanishes everywhere. Thus the spacetime is
vacuum. This in turn implies that $m$ is zero on the whole spacetime.
It follows that $\theta$ and $\theta'$ are identically zero and that
$r$ is constant. In vacuum the Gaussian
curvature of the metric $g_{ab}$ is $K=r^{-1}\Delta r$ and so
in the present case $K=0$ and $g_{ab}$ is flat. It is easily
seen that this and the constancy of $r$ imply that the spacetime
is flat.

\vskip 10pt
In the case of hyperbolic symmetry the following analogue of Lemma 2.4
holds.

\noindent
{\bf Lemma 2.5} Let $(M,g)$ be a spacetime with hyperbolic symmetry
which satisfies the dominant energy condition. Then $\nabla^a r$ is
timelike.

\noindent
{\bf Proof} $\nabla^a r$ must be timelike somewhere on a symmetric
Cauchy surface, as shown in the proof of Lemma 2.3. Hence it suffices
to show that $\nabla_a r\nabla^a r$ never vanishes. This follows
immediately from the analogue of (2.21).

\vskip 10pt
Non-flat spacetimes with plane symmetry and all spacetimes with
hyperbolic symmetry have the property that the gradient of $r$ is
either everywhere past-pointing timelike or everywhere future-pointing
timelike. Spacetimes where the first possibility is realized may
be called \lq expanding\rq\ models those where the second is realized
\lq contracting\rq\ models. This terminology is justified by the
following considerations. If the gradient of $r$ is past pointing
then $\theta$ is positive and $\theta'$ is negative. From (2.10) then
follows that $t<K$ everywhere. If $t$ were non-negative then this would
mean that $|K|$ was everywhere greater than $|t|$. However this is
inconsistent with the existence of a compact CMC hypersurface, as can
be seen by integrating the Hamiltonian constraint (2.3). Thus if the
gradient of $r$ is past-pointing in a region foliated by compact
CMC hypersurfaces then $t$ must be negative. Similarly, if the
gradient of $r$ is future-pointing, $t$ must be positive. By possibly
replacing $t$ by $-t$ it can be assumed without loss of generality
that the model is expanding i.e. that the $r$ increases monotonically
with $t$ along any causal curve. In that case $t<0$ in the whole
spacetime so that $t_1\le 0$. In this case it will be assumed that in
fact $t_2<0$.

The information obtained so far in this section implies bounds for
various geometrical quantities in a surface symmetric spacetime
without centre defined on a finite
time interval $(t_1,t_2)$. Now as many other quantities as possible
will be bounded.  Suppose that the energy-momentum tensor
satisfies the non-negative pressures condition. Then (2.17) implies
that when $\nabla_a r$ is timelike the rates of change of $r$ and $m$
along an integral curve of $\nabla_a r$ have opposite signs. It
follows that in the cases $\epsilon=0$ and $\epsilon=-1$ the radius
$r$ is bounded below by a positive constant and the mass bounded above
on the interval $(t_3,t_2)$ while the radius is bounded above and the
mass bounded below on the interval $(t_1,t_3)$, where $t_3$ is any time
with $t_1<t_3<t_2$. If $\epsilon=0$ or if $\epsilon=-1$ and
it is assumed that $m$ is
positive for $t=t_3$ then the mass is bounded below by a positive
constant on $(t_1,t_3)$. In the case
$\epsilon=1$ results of Burnett[2] show that the radius is bounded
above and the mass bounded below by a positive constant on both these
intervals. Using the upper bound for $2m/r$ obtained earlier it can be
seen that on an interval where the radius is bounded above and the
mass bounded below by a positive constant the mass is bounded above and
the radius is bounded away from zero.

An estimate for the lapse function $\alpha$ can be obtained from
(2.4). Considering a point where $\alpha$ attains its maximum and
using the fact that $\rho+\tr S\ge 0$
shows that $\alpha\le 3/t^2$. Hence if $t_1\ne0$ and
$t_2\ne0$ it follows that $\alpha$ is bounded on $(t_1,t_2)$.
An interval which satisfies $t_1\ne0$ and $t_2\ne0$ and where $r$ is
bounded above and $m$ is bounded below by a positive constant
will be called admissible. The volume of the slice $t={\rm const.}$ is
$V(t)=C\int_0^{2\pi}a^2A^3$ and its time evolution is given by
$$dV/dt=-Ct \int_0^{2\pi}\alpha a^2A^3\eqno(2.22)$$
In the case of an admissible interval this shows that $V(t)$ and its
inverse are bounded.
Now $V=a^{-1}\int_0^{2\pi}r^3$ and so if $t_1\ne0$, $t_2\ne0$
there are positive constants $C_1$ and $C_2$ such that $C_1\le a\le C_2$
and $C_1\le A\le C_2$.
Now integrate the equation (2.3) from 0 to
$2\pi$. There results the inequality
$$\int_0^{2\pi}16\pi\rho A^{5/2}\le 2\int_0^{2\pi}A^{-1/2}a^{-2}
+\f23\int_0^{2\pi}t^2 A^{5/2}\eqno(2.23)$$
which implies that on an admissible interval $\int_0^{2\pi}\rho$ is
bounded. The dominant energy condition then gives bounds for $\int j$
and $\int\tr S$. The bounds for $r\theta$ and $r\theta'$ now show that
$A'$ and $K$ are bounded on any admissible time interval. Integrating
(2.6) over the circle allows $a_t$ to be bounded and then (2.6) itself
gives a bound for $\beta'$. A bound for $\beta$ can be deduced using
the condition $\int\beta=0$. Equation (2.8) gives a bound for $A_t$.
Integrating equation (2.4) from a point where $\alpha'=0$ and using
(2.23) provides a bound for $\alpha'$.

\noindent
{\bf Theorem 2.1} Let a solution of the Einstein equations with surface
symmetry be given and suppose that when coordinates are chosen which
cast the metric into the form (2.1) with constant mean curvature time
slices the time coordinate takes all values in the finite interval
$(t_1,t_2)$. Suppose further that:
\next
i) the dominant energy and non-negative pressures conditions hold
\next
ii) neither $t_1$ or $t_2$ is zero
\next
iii) if $\epsilon$ is $0$ or $1$ then $t_1<0$
\next
iv) if $\epsilon=-1$ then the mass function is positive on the initial
hypersurface.
\next
Let $t_3$ satisfy $t_1<t_3<t_2$. Then the following
quantities are bounded on $(t_1,t_3)$:
$$\eqalignno{
&\alpha,\alpha',A,A^{-1},A',K,\beta,a,a^{-1},\d_t a&(2.24)      \cr
&\d_t A,K',\beta'&(2.25)}$$

\vfil\eject

\noindent
{\bf 3. The matter fields}

Theorem 2.1 provides some information on the boundedness of certain
geometrical quantities in a spacetime with spherical, plane or
hyperbolic symmetry without any assumptions on the matter content
except the dominant energy and non-negative pressures conditions. To
get further bounds and hence to proceed towards showing that the
spacetime can be extended it is necessary to use the matter field
equations. In the following two examples will be treated, namely the
collisionless gas and the massless scalar field.

The collisionless gas is described by a distribution function $f$
which is a non-negative real-valued function on the mass shell.
It is supposed to satisfy the Vlasov equation which in the class
of spacetimes considered here takes the form:
$$\eqalign{
{\d f\over\d t}&+\left(\alpha A^{-1}{v^1\over v^0}-\beta\right)
{\d f\over\d x}+\left[-A^{-1}\alpha'v^0+\alpha Kv^1+\alpha A^{-2}A'
{(v^2)^2+(v^3)^2\over v^0}\right]{\d f\over\d v^1}              \cr
&\qquad -\alpha\left[A^{-2}A'{v^1\over v^0}+{1\over 2}\left(K-t
\right)\right]v^B{\d f\over\d v^B}=0}\eqno(3.1)$$
Here the mass shell has been coordinatized using components in an
orthonormal frame, where the first vector in the spatial frame is
proportional to $\d/\d x$. The component $v^0$ is then given by
the expression $\sqrt{1+(v^1)^2+(v^2)^2+(v^3)^2}$.
The upper case Latin indices take the values $2$ and $3$. The
distribution function depends on $t$, $x$, $v^1$, $v^2$ and $v^3$.
In fact the symmetry requires that its dependence on the last two
quantities is only a dependence on the combination $(v^2)^2+(v^3)^2$.
This will be assumed for the initial data and is then also satisfied
by the solution. The initial data is assumed to be compactly
supported and then the solution has compact support at each fixed
time. That the initial value problem for the Vlasov-Einstein system
is well posed was shown by Choquet-Bruhat[3]. The matter quantities
occurring in the field equations are given by
$$\eqalign{
\rho&=\int fv^0 dv             \cr
j&=\int fv^1 dv                  \cr
S^1_1&=\int f(v^1)^2/v^0 dv       \cr
\tr S&=\int f [(v^0)^2-1]/v^0 dv }
\eqno(3.2)$$
For a solution which evolves from initial data given at $t=t_0$ let
$$P(t)=\sup\{|v|:f(s,x,v)\ne 0\ {\rm for\ some}\ (s,x,v)\ {\rm with}\
s\in I\}\eqno(3.3)$$
where $I$ is the interval $[t_0,t]$ if $t\ge t_0$ and the interval
$[t,t_0]$ if $t\le t_0$.
The maximum of $f$ is time independent and so all the matter
quantites defined in (3.2) can be bounded by $C(1+P(t))^4$.
The quantity $P(t)$ itself can be controlled by studying the
characteristics of the equation (3.1) since this equation
says that $f$ is constant along these characteristics. It
follows that $P(t)$ will be bounded on a given interval provided
the coefficients in (3.1) are bounded. The geometrical quantities
which occur in these coefficients are $\alpha$, $A^{-1}$, $\beta$,
$K$ and $t$. Theorem 2.1 shows that all of these are bounded on
an admissible time interval. Considering a point where $\alpha$
attains its minimum on a given hypersurface $t$=const. leads to
a bound for $\alpha^{-1}$. Hence the following is obtained.

\noindent
{\bf Theorem 3.1} Let a solution of the Vlasov-Einstein system
with surface symmetry be given and suppose that
when coordinates are chosen which cast the metric into the form (2.1)
with constant mean curvature time slices the time coordinate takes
all values in the finite interval $(t_1,t_2)$. Suppose further that
conditions (ii)-(iv) of Theorem 2.1 are satisfied. Then
all the quantities in (2.24) and (2.25) are bounded $(t_1,t_3)$, as are
$P$, $\alpha^{-1}$ and
$$\rho, j, S^1_1,\tr S\eqno(3.4)$$

Notice that while Theorem 2.1 is to a large extent independent of
the matter model used this is not the case for Theorem 3.1. It
is probable that the analogous statement would be false if the
collisionless matter was replaced by dust since shell-crossing
singularities would presumably provide counterexamples.
It will now be shown by induction that under the hypotheses of
Theorem 3.1 all derivatives of the solution are bounded.

\noindent
{\bf Lemma 3.1} If the hypotheses of Theorem 2.1 are satisfied
and if all derivatives with respect to $x$ of order up to $n$
of the quantities in (2.24) and (3.4) are bounded then
all derivatives with respect to $x$ of order up to $n+1$ of the
quantities in (2.24) are bounded.

\noindent
{\bf Proof} In the following $D_x$ denotes a derivative with
respect to $x$. Note first that the boundedness of the derivatives
with respect to $x$ up to order $n+1$ of the quantities $\alpha$,
$A$, $A^{-1}$, $a$ and $a^{-1}$ follows immediately from the
hypotheses of the lemma. Equations (2.3)-(2.6) can be solved
for the quantities $A''$, $\alpha''$, $K'$ and $\beta'$. Differentiating
the resulting equations $n$ times with respect to $x$ allows
$D_x^{n+1}(A')$, $D_x^{n+1}(\alpha')$, $D_x^{n+1}K$ and $D_x^{n+1}\beta$
to be bounded.

\vskip 10pt
\noindent
{\bf Lemma 3.2} If the hypotheses of Lemma 3.1 are satisfied
by a solution of the Vlasov-Einstein system then all derivatives
with respect to $x$ of order up to $n+1$ of the quantities in (3.4)
are bounded. Moreover all derivatives of $f$ of order up to $n+1$
with respect to $x$ and $v$ are bounded.

\noindent
{\bf Proof} The hypotheses imply that the coefficients of the
characteristic system are bounded together with their derivatives
with respect to $x$ up to order $n+1$. Differentiating this system
$n+1$ times with respect to $x$ and $v$ gives an inhomogeneous linear
system of ordinary differential equations for the derivatives of order
$n+1$ of the unknowns. The coefficients of this system are bounded,
as long as attention is confined to the support of $f$. Hence these
derivatives are bounded on the support of $f$. It follows from this that
the derivatives of $f$ up to order $n+1$ are bounded. Differentiating
(3.2) then gives the desired conclusion for the quantities in (3.4).

\vskip 10pt
\noindent
{\bf Lemma 3.3} If the hypotheses of Theorem 2.1 are satisfied
by a solution of the Vlasov-Einstein system and if all derivatives
of the quantities in (2.24) and (3.4) of the form $D_t^kD_x^n$
with $n$ arbitrary and $k\le m$ are bounded and if all higher
derivatives of $f$ with at most $m$ time derivatives are bounded
then the derivatives of the form $D_t^{m+1}D_x^n$ of the quantities in
(3.4) are bounded. Moreover the higher derivatives of $f$ with at
most $m+1$ time derivatives are bounded.

\noindent
{\bf Proof} Use the Vlasov equation to bound an extra time derivative
of $f$ and substitute the result into (3.2).

\vskip 10pt\noindent
{\bf Lemma 3.4} If the hypotheses of Lemma 3.1 are satisfied and if:
\next
(i) all derivatives of the quantities in (2.24) and (3.4) of the
form $D_t^kD_x^n$ with $n$ arbitrary and $k\le m$ are bounded
\next
(ii) all derivatives of the quantities in (3.4) of the form
$D_t^{m+1}D_x^n$ are bounded

\noindent
then the derivatives of the quantities in (2.24) of the form
$D_t^{m+1}D_x^n$ are bounded.

\noindent
{\bf Proof} The conclusion for $a$ and $a^{-1}$ follows immediately
from the assumptions. The conclusion for $A$, $A^{-1}$, $A'$ and $K$
follows from equations (2.8) and (2.9). Now differentiate (2.4)
$n$ times with respect to $x$ and $m+1$ times with respect to $t$.
The result is an equation of the form
$$\eqalign{
&(D_t^{m+1}D_x^n\alpha)''+A^{-1}A'(D_t^{m+1}D_x^n\alpha)'     \cr
&\qquad=A^2[\f32(K-\f13t)^2+\f13 t^2+4\pi(\rho+\tr S)]D_t^{m+1}D_x^n\alpha
+B}$$
where the remainder term $B$ is bounded. Examining the points where
$D_t^{m+1}D_x^n\alpha$ has maximum modulus gives
$|D_t^{m+1}D_x^n\alpha|\le 3A^{-2}B/t^2$. Next, the equation (2.7),
integrated with respect to $x$, allows the conclusion to be obtained
for $\d_t a$. Finally, the conclusion for $\beta$ follows from (2.6).

\vskip 10pt\noindent
Putting together the conclusions of Theorem 3.1 and Lemma 3.1-3.4
we see that under the hypotheses of Theorem 3.1 all derivatives of all
metric coefficients and of the distribution function are bounded.

When all the derivatives of the solution are bounded on a given
interval, it can be extended smoothly to the closure of that interval.
There results a new initial data set and applying the local
existence and uniqueness result discussed in Appendix 1 gives an
extension of the solution to an interval which strictly contains
the original one. Thus we have the following theorem.

\noindent
{\bf Theorem 3.2} Let $(M,g,f)$ be a $C^\infty$ solution of the
Vlasov-Einstein system with surface symmetry which is the maximal
globally hyperbolic development of data given on a hypersurface of
constant mean curvature $H_0$. Then:
\next
1. If $\epsilon=1$ and $H_0=0$ then the whole spacetime can be
covered by a foliation of CMC hypersurfaces where the mean curvature
takes all real values.
\next
2. If $\epsilon=1$ or $\epsilon=0$ and $H_0<0$ then the part of the
spacetime to the past of the initial hypersurface can be covered by
a foliation of CMC hypersurfaces where the mean curvature takes all
values in the interval $(-\infty,H_0]$
\next
3. If $\epsilon=-1$, $H_0<0$ and the mass function is positive on
the initial hypersurface then the part of the spacetime to the past
of the initial hypersurface can be covered by a foliation of CMC
hypersurfaces where the mean curvature takes all values in the
interval $(-\infty,H_0]$

Now another example, the massless scalar field, will be discussed.
The massless scalar field is described by a real-valued function
$\phi$ satisfying $\nabla_\alpha\nabla^\alpha\phi=0$. The
energy-momentum tensor is
$$T_{\alpha\beta}=\nabla_\alpha\phi\nabla_\beta\phi-\f12
(\nabla_\gamma\phi\nabla^\gamma\phi)g_{\alpha\beta}\eqno(3.5)$$
The dominant energy condition is satisfied but the non-negative
pressures condition does not hold for a general $\phi$. Hence
Theorem 2.1 does not apply to this case. However,
as has been remarked by Burnett[2], it is true that $T_{\alpha\beta}
x^\alpha x^\beta\ge0$ for spacelike vectors $x^\alpha$ orthogonal to
the surfaces of symmetry and it turns out that this fact and the
condition that $\rho+\tr S\ge 0$ are the only ones which are
needed in the proof of the theorem. They are satisfied by the
scalar field (see below). There is also a potential
problem with applying the analysis of Appendix 1 in this case,
since there the non-negative pressures condition was also used.
It was needed for the argument using the implicit function theorem
but only in the case that the initial hypersurface is maximal.
Hence the analogue of Theorem 2.1 holds for the scalar field
under the extra hypothesis that the interval $(t_1,t_2)$ does
not contain zero.

To proceed further it is useful to introduce the null vectors
$$\eqalign{
e_+&=\alpha^{-1}\left({\d\over\d t}-\beta{\d\over\d x}\right)
+A^{-1}{\d\over \d x}
\cr
e_-&=\alpha^{-1}\left({\d\over\d t}-\beta{\d\over\d x}\right)
-A^{-1}{\d\over \d x}
}\eqno(3.6)$$
Let $\phi_+=e_+\phi$ and $\phi_-=e_-\phi$. Then the wave equation
for $\phi$ can be written as
$$\eqalignno{
\alpha e_+(\phi_-)&=\alpha t(\phi_++\phi_-)
+(A'\alpha'+2\alpha A^{-2}A')(\phi_+-\phi_-)+\alpha[e_-,e_+]\phi
&(3.7)   \cr
\alpha e_-(\phi_+)&=\alpha t(\phi_++\phi_-)
+(A'\alpha'+2\alpha A^{-2}A')(\phi_+-\phi_-)+\alpha[e_+,e_-]\phi
&(3.8)   \cr
}$$
Now $\alpha [e_+,e_-]=b_+e_++b_-e_-$, where the coefficients $b_+$
and $b_-$ are polynomials in the quantities (2.24) and (2.25).
In addition, the definitions of $e_+$ and $e_-$ imply that
$$\d\phi/\d t-\beta\d\phi/\d x=\f12\alpha(e_++e_-)\eqno(3.9)$$
The matter quantities occurring in the field equations are given by
$$\eqalign{
\rho&=\f14(\phi_+^2+\phi_-^2)                      \cr
j&=\f14(-\phi_+^2+\phi_-^2)                        \cr
S^1_1&=\f14(\phi_+^2+\phi_-^2)                     \cr
\tr S&=\f14(\phi_++\phi_-)^2+\f12\phi_+\phi_-      \cr
\rho+\tr S&=\f12(\phi_++\phi_-)^2}
\eqno(3.10)$$
Let
$$\Phi(t)=\|\phi(t)\|_\infty+\|\phi_+(t)\|_\infty+\|\phi_-(t)\|_\infty
\eqno(3.11)$$
The equations (3.7)-(3.9) and the boundedness of the quantities
(2.24) and (2.25) imply that
$$\Phi(t)\le \Phi(0)+C\int_0^t \Phi(s) ds\eqno(3.12)$$
Hence $\phi$, $\phi_+$ and $\phi_-$ are bounded on an admissible
interval. It then follows from (3.10) that the matter quantities
in the field equations are bounded. Hence the analogue of Theorem
3.1 with the Vlasov-Einstein system replaced by the Einstein-scalar
system and $P$ replaced by $\Phi$ holds provided the interval
$(t_1,t_2)$ does not contain zero.

\vskip 10pt
\noindent
{\bf Lemma 3.5} If the hypotheses of Lemma 3.1 are satisfied
by a solution of the Einstein-scalar system then all derivatives
with respect to $x$ of order up to $n+1$ of the quantities in (3.4)
are bounded. Moreover all derivatives of $\phi$ of order up to $n+1$
and of $\d_t\phi$ up to order $n$ with respect to $x$ are bounded.

\noindent
{\bf Proof} Note first that the results of Lemma 3.1 can be improved
slightly by using (2.6) and (2.8) to bound the spatial derivatives of
order up to order $n+1$ of the quantities $\beta'$ and $\d_t A$.
Next, differentiating the equations (3.7)-(3.9) gives an inhomogeneous
linear hyperbolic system for $D_x^{n+1}\phi_+$, $D_x^{n+1}\phi_-$ and
$D_x^{n+1}\phi$ with a bounded right hand side. This gives the desired
bounds for derivatives of $\phi$. Putting these into (3.10) gives
the bounds for the components of the energy-momentum tensor.

\vskip 10pt
\noindent
{\bf Lemma 3.6} If the hypotheses of Theorem 2.1 are satisfied
by a solution of the Einstein-scalar system and if all derivatives
of the quantities in (2.24) and (3.4) of the form $D_t^kD_x^n$
with $n$ arbitrary and $k\le m$ are bounded and if all higher
derivatives of $\phi$ with at most $m+1$ time derivatives are bounded
then the derivatives of the form $D_t^{m+1}D_x^n$ of the quantities in
(3.4) are bounded. Moreover the higher derivatives of $\phi$ with at
most $m+2$ time derivatives are bounded.

\noindent
{\bf Proof} This result follows immediately from the equations
(3.7)-(3.9) and (3.10)

Putting together the analogue of Theorem 3.1 for the scalar field
and the Lemmas 3.1, 3.4, 3.5 and 3.6, we see that under the hypotheses
of the analogue of Theorem 3.1 and assuming that $H_0\ne 0$ all derivatives
of all metric coefficients and of the scalar field are bounded.
If $H_0=0$ then we still get a neighbourhood of the initial hypersurface
foliated by CMC hypersurfaces. If there exists a hypersurface belonging
to this foliation with non-zero mean curvature both to the past and to
the future of the initial hypersurface then the problem of getting bounds
is reduced to the case $H_0\ne 0$. If all the CMC hypersurfaces belonging
to the foliation which are to the past, say, of the initial hypersurface
are maximal then the second fundamental form and the Ricci tensor
contracted twice with the normal vector are both zero on that region.
The latter implies that the gradient of $\phi$ is tangent to the CMC
hypersurfaces. Under these circumstances the wave equation reduces
to the Laplace equation and $\phi$ must be spatially constant. If
$\phi$ varied from one spacelike hypersurface to the next then $\nabla
^\alpha\phi$ would be timelike, contradicting what has been said already.
Hence $\phi$ is constant, the energy-momentum tensor is zero and the
vacuum Einstein equations are satisfied. These considerations show that the
following analogue of Theorem 3.2 holds for the scalar field:

\noindent
{\bf Theorem 3.3} Let $(M,g,\phi)$ be a $C^\infty$ solution with surface
symmetry of the Einstein equations coupled to a massless scalar field
which is the maximal globally hyperbolic development of data given on
a hypersurface of constant mean curvature $H_0$. Then:
\next
1. If $\epsilon=1$ and $H_0=0$ then the whole spacetime can be
covered by a foliation of CMC hypersurfaces where the mean curvature
takes all real values.
\next
2. If $\epsilon=1$ or $\epsilon=0$ and $H_0<0$ then the part of the
spacetime to the past of the initial hypersurface can be covered by
a foliation of CMC hypersurfaces where the mean curvature takes all
values in the interval $(-\infty,H_0]$
\next
3. If $\epsilon=-1$, $H_0<0$ and the mass function is positive on
the initial hypersurface then the part of the spacetime to the past
of the initial hypersurface can be covered by a foliation of CMC
hypersurfaces where the mean curvature takes all values in the
interval $(-\infty,H_0]$

\vskip .5cm\noindent
{\bf Appendix 1}

The purpose of this appendix is to prove Lemma 2.1 and a local
existence and uniqueness theorem for the equations (2.3)-(2.9). Let
$(M,g_{\alpha\beta})$ be a surface symmetric spacetime,
as defined in Section 2. Thus, in particular $M$ is of the form
$\R\times S^1\times F$. Let $\tilde M$ be the universal cover of $M$
and let $\tilde g_{\alpha\beta}$ the pull-back of $g_{\alpha\beta}$ to
$\tilde M$. Suppose that $(M,g_{\alpha\beta})$ contains a CMC Cauchy
hypersurface $S$. It follows from the fact[10] that all globally defined
Killing vectors must be tangent to a compact CMC hypersurface that a
compact CMC Cauchy hypersurface $S$ in a spacetime with spherical
or plane symmetry is of the form $\bar S\times F$. For in that case
there are enough global Killing vectors to generate the surfaces of
symmetry. This is not true in the case of hyperbolic symmetry.
It is a standard fact that a neighbourhood of a compact CMC
hypersurface can be foliated by compact CMC hypersurfaces unless the
original hypersurface is such that its second fundamental form vanishes
and the Ricci tensor contracted twice with the normal vector is zero.
Furthermore the mean curvature of these hypersurfaces can be used as a
time coordinate in this neighbourhood. Even if this condition fails
there is still a neighbourhood of the initial hypersurface foliated
by CMC hypersurfaces, although in that case the mean curvature cannot
be used as a time coordinate[1]. When the dominant energy
and non-negative pressures conditions are satisfied the condition
can only fail if the spacetime is vacuum on $S$. From what was said
earlier the CMC hypersurfaces must be symmetric in the cases of
spherical and plane symmetry. In fact they must also be symmetric
in the hyperbolic case. To see this note that the
existence of these hypersurfaces is proved by using the inverse
function theorem. However it is possible to apply the inverse
function theorem in the class of symmetric deformations of the
initial hypersurface (provided this initial hypersurface is itself
symmetric) and then the CMC hypersurfaces obtained are by construction
symmetric. Consider now the exceptional case where the spacetime is
vacuum and the second fundamental form of the initial hypersurface
vanishes. In the cases of spherical
and hyperbolic symmetry integrating (2.3) from $0$ to $2\pi$ gives a
contradiction. In the plane symmetric case the vanishing of the
second fundamental form implies that spacetime is flat.
Hence unless the spacetime is flat the mean curvature can be
used as a time coordinate $t$ in a neighbourhood of $S$. The inverse image
of $t$ under the projection $p:\tilde M\to M$ will also be denoted by $t$.

It is elementary to see that the metric $g_{ab}$ of the hypersurface
$t={\rm const.}$ in $M$ can be written locally in the form
$$A^2 dx^2+B^2 d\Sigma^2\eqno(A1.1)$$
where $d\Sigma^2$ is a metric of constant curvature. What is less clear is
that that this can be done globally in such a way that $x\in [0,2\pi]$
and the functions $A$ and $B$ are $2\pi$-periodic. Consider one of the
hypersurfaces $t={\rm const.}$ in $\tilde M$. Let $\gamma$ be a
geodesic in this hypersurface which starts orthogonal to one of the
group orbits ${\cal O}_1$ at a point $p$. It continues to be
orthogonal to the orbits. After a finite time it must hit an orbit
${\cal O}_2$ which projects to the same orbit in $M$ as ${\cal O}_1$.
Suppose it meets ${\cal O}_2$ at a point $q$. Let $q'$ be the unique
point of ${\cal O}_1$ which projects to the same point of $M$ as
$q$. Any isometry which fixes $p$ must fix $q'$. It follows in the
plane and hyperbolic cases that $p=q'$ and in the spherical case that
either $p=q'$ or $p$ and $q'$ are antipodal points on the sphere. If $p=q'$
then $A$ and $B$ can be made $2\pi$-periodic, as desired. In the
case where the antipodal map occurs the same thing can be arranged
by allowing $x$ to go twice around the circle.

Let $$a=2\pi\left[\int_0^{2\pi}B(x)/A(x)dx\right]^{-1}\eqno(A1.2)$$
Then the new coordinate $x'$ defined by
$$x'=a\int_0^x B(x)/A(x)dx\eqno(A1.3)$$
satisfies $x(0)=0$ and $x(2\pi)=2\pi$. Define
$A'$ by the relation $A'(x'){dx'\over dx}(x)=A(x)$. After
transforming to the new coordinate and dropping the primes the metric
takes the form
$$A^2(dx^2+a^2 d\Sigma^2)\eqno(A1.4)$$
where $A$ is a positive function of $x$ with $A(0)=A(2\pi)$ and $a$ is a
constant. Doing this construction on each hypersurface of constant
time gives the coordinate system whose existence is asserted by Lemma
2.1.

Consider now the question of local existence and uniqueness of
solutions of equations (2.3)-(2.9) with given initial data on
a hypersurface $t={\rm const.}$. In order to have a well-posed
initial value problem it is necessary to have some matter equations
such that the resulting Einstein-matter system has a well-posed
Cauchy problem in the context of $C^\infty$ data and solutions.
More precisely we assume that the solution of the initial value
problem for the reduced equations in harmonic coordinates exists
and is unique so that the general theory of the maximal Cauchy
development [4] can be applied. (If it were desired to consider
gauge theories, where solutions are only unique up to gauge
transformations, then some more work would be required.) It
is not obvious that this theory applies to kinetic theory models,
where the matter fields are defined on the mass shell rather
than on spacetime. However the analogous results do hold in that
case [3].
An initial data set consists of periodic functions $A$ and $K$,
a constant $a$ and matter data which satisfy the constraint
equations (2.3) and (2.5). The matter data are assumed to have
the necessary symmetry properties. These properties are most
easily expressed on the covering manifold. The data set on the
covering manifold has a maximal Cauchy development on the manifold
$\tilde M$. The maximal Cauchy development inherits the symmetries
of the data and so the original initial data set has a surface
symmetric Cauchy development. In this surface symmetric spacetime
coordinates can be introduced as above. Thus a solution of equations
(2.3)-(2.9) (and the matter equations) on some interval $(t_1,t_2)$
is obtained. It remains to show that solutions of these equations
are uniquely determined by initial data. Suppose there exist two
solutions with the same initial data on the interval $(t_1,t_2)$.
Then by the general theory of the Cauchy problem there must exist
embeddings $\phi_1$ and $\phi_2$ of $M$ into the maximal Cauchy
development of the given initial data set such that $\phi_1$ is
a matter preserving isometry for the first solution and $\phi_2$
a matter preserving isometry for the second. The uniqueness of
compact constant mean curvature hypersurfaces implies that the
images of any hypersurface of constant $t$ under $\phi_1$ and
$\phi_2$ are identical. In particular this means that the images
of $M$ under $\phi_1$ and $\phi_2$ are identical, so that there
exists a diffeomorphism $\phi_{12}:M\to M$ such that $\phi_1=
\phi_{12}\circ\phi_2$. The diffeomorphism $\phi_{12}$ maps the
one solution into the other and preserves the hypersurfaces of
constant time. Suppose temporarily that the initial data do not
have Robertson-Walker symmetry. Then a unique two-plane is defined
by the isotropy group of the universal cover. This plane must be
preserved by $\phi_{12}$ as must its orthogonal complement. Thus, if
$\phi_{12}$ is written in terms of coordinates adapted to the first
metric in the form $(t,x,y)\mapsto (t',x',y')$. Then $t'=t$ and $y'$
depends only on $t$ and $y$. By composing with an isometry it can be
reduced to the identity on the initial hypersurface. The form
of the shift vector then implies that it is the identity everywhere.
In a given spacetime the coordinate $x$ is defined up to a
translation. Hence $x'=x+c$ and since $\phi_{12}$ is the identity
on the initial hypersurface it follows that the two solutions are
identical. In the case where the data have Robertson-Walker symmetry
the solutions must have Robertson-Walker symmetry and in that case
uniqueness for the reduced equations is obvious.

\vskip.5cm\noindent
{\bf Appendix 2}

In this appendix the vacuum solutions with the symmetry properties
considered in this paper will be determined. This is done using
the following lemma:

\noindent
{\bf Lemma A2.1} Consider the ordinary differential equation
$d^2 u/dx^2=f(u)$, where $f:(0,\infty)\to{\bf R}$ is Lipschitz.
Suppose that $f(u_0)=0$ for some $u_0$, $f(u)<0$ for $0<u<u_0$
and $f(u)>0$ for $u>u_0$. Then any periodic solution is constant.

\noindent
{\bf Proof} Let $u$ be a periodic solution. By periodicity there
exists a point $x_0$ where $d^2 u/dx^2$ vanishes. At that point
$u=u_0$. If $du/dx (x_0)$ is positive then it is easy to show
that $du/dx$ remains positive for $x>x_0$, contradicting
periodicity. Similarly the assumption $du/dx (x_0)<0$ leads
to a contradiction. Hence in fact $du/dx(x_0)=0$. By uniqueness
for solutions of the ordinary differential equation it follows
that $u$ is constant.

\vskip 10pt
Consider now vacuum solutions of equations (2.3)-(2.9). The momentum
constraint can be solved explicitly, giving $K-\f13 t=CA^{-3}$ for
some constant $C$. Substituting this into the Hamiltonian constraint
gives:
$$(A^{1/2})''=-\f3{16}C^2A^{-7/2}+\f1{12}t^2A^{5/2}+\f14\epsilon
a^{-2}A^{-1/2}\eqno(A2.1)$$
It can be checked straightforwardly that this ordinary differential
equation for $A^{1/2}$ satifies the hypotheses of the lemma and so
$A$ is constant. Then the same lemma may be applied to the lapse
equation to show that $\alpha$ is constant. The constancy of $A$
implies that of $K$ and the equation for $\beta$ then gives $\beta=0$.
Hence every vacuum solution of equations (2.3)-(2.9) is spatially
homogeneous. These solutions will now be identified with known
exact solutions. This will be done by examining the Cauchy data
on one spacelike hypersuface. Suppose that constants $t$, $a$ and
$K$ are given and satisfy the following sign condition, which is
necessary for the constraints to have a solution:

\noindent
(i) if $\epsilon=1$ then $\f32(K-\f13 t)^2-\f23 t^2>0$

\noindent
(ii) if $\epsilon=0$ then $\f32(K-\f13 t)^2-\f23 t^2=0$

\noindent
(iii) if $\epsilon=-1$ then $\f32(K-\f13 t)^2-\f23 t^2<0$

Suppose that $t=0$. Then the sign condition is incompatible with
$\epsilon=-1$. It is only compatible with $\epsilon=0$ if $K=0$.
In that case the data give rise to flat space, identified in a
simple way. If $\epsilon=1$ then the Hamiltonian constraint can
be solved for $A$ in terms of $a$ and $K$. For $t\ne 0$ the sign
condition can readily be analysed by dividing the expression of
interest by $t^2$ and studying the resulting quadratic expression
in $K/t$.

The case $\epsilon=0$ is the simplest. There are two possible
values for $K/t$, namely $-1/3$ and $1$. These solutions of
the constraints can be realized by the $\tau$=const. hypersurfaces
in the Kasner solution
$$-d\tau^2+b^2\tau^{2p}dx^2+\tau^{1-p}(dy^2+dz^2)\eqno(A2.2)$$
where $p=-1/3$ or $p=1$ and $b$ is a positive constant. In the
case $\epsilon=1$ the quantity $K/t$ takes all values in the
intervals $(-\infty,-1/3)$ and $(1,\infty)$ and these solutions
of the constraints can be realized by the $\tau$=const. hypersurfaces
in the following metric, which is obtained by identifying the part
of the Schwarzschild solution inside the horizon:
$$-(2m/\tau-1)^{-1}d\tau^2+b^2(2m/\tau-1)dx^2+\tau^2 d\Sigma^2\eqno(A2.3)$$
Here $d\Sigma^2$ is the standard metric on the sphere.
Similarly, the solutions with $\epsilon=-1$ produce all values of
$K/t$ in the interval $(-1/3, 1)$ and these solutions of the constraints
can be realized by the $\tau$=const. hypersurfaces in the
following pseudo-Schwarzschild metric:
$$-(2m/\tau+1)^{-1}d\tau^2+b^2(2m/\tau+1)dx^2+\tau^2 d\Sigma^2\eqno(A2.4)$$
In this case $d\Sigma^2$ is a metric of constant negative curvature
on a compact manifold obtained by identifying the hyperbolic plane
by means of a discrete group of isometries.
The general theorems proved in this paper imply in particular that for
$m>0$ the initial singularity in this solution, which occurs at $t=0$
is a crushing singularity. It is worth remarking that for $m\le 0$
the initial singularity, which occurs at $t=-2m$ is also crushing, even
through the theorems do not apply.

\vskip .5cm\noindent
{\bf References}

\noindent
1. Bartnik, R.: Remarks on cosmological spacetimes and constant mean
curvature hypersurfaces. Commun. Math. Phys. 117, 615-624 (1988).
\next
2. Burnett, G.: Incompleteness theorems for the spherically
symmetric spacetimes. Phys. Rev. D43, 1143-1149 (1991).
\next
3. Choquet-Bruhat, Y.: Probl\`eme de Cauchy pour le syst\`eme
integro differentiel d'Einstein-Liouville. Ann. Inst. Fourier
21, 181-201 (1971).
\next
4. Choquet-Bruhat, Y., Geroch, R.: Global aspects of the Cauchy problem
in general relativity. Commun. Math. Phys. 14, 329-335 (1969).
\next
5. Chru\'sciel, P. T.: On spacetimes with $U(1)\times U(1)$ symmetric
compact Cauchy surfaces. Ann. Phys. 202, 100-150 (1990).
\next
6. Eardley, D., Moncrief, V.: The global existence problem and cosmic
censorship in general relativity. Gen. Rel. Grav. 13, 887-892 (1981).
\next
7. Eardley, D., Smarr, L.: Time functions in numerical
relativity: marginally bound dust collapse. Phys. Rev. D19,
2239-2259 (1979).
\next
8. Hartman, P.: Ordinary Differential Equations. (Birkh\"auser, 1982).
\next
9. Holcomb, K.: A computer code for the study of spherically symmetric
cosmological spacetimes. Gen. Rel. Grav. 22, 145 (1990).
\next
10. Isenberg, J., Moncrief, V.: The existence of constant mean curvature
foliations of Gowdy 3-torus spacetimes. Commun. Math. Phys. 86,
485-493 (1983).
\next
11. Malec, E., \'O Murchadha, N.: Optical scalars in spherical
spacetimes. Preprint gr-qc/9404031.
\next
12. Rein, G.: Cosmological solutions of the Vlasov-Einstein system with
spherical, plane and hyperbolic symmetry. Preprint gr-qc/9409041.
\next
13. Rendall, A. D.: Cosmic censorship for some spatially homogeneous
cosmological models. Ann. Phys. 233, 82-96 (1994).
\next
14. Rendall, A. D.: On the nature of singularities in plane symmetric
scalar field cosmologies. Preprint gr-qc/9408001.
\next
15. Rendall, A. D.: Global properties of locally spatially homogeneous
cosmological models with matter. Preprint gr-qc/9409009, to appear in
Math. Proc. Camb. Phil. Soc.
\end